\documentclass[11pt,a4paper]{article}

\usepackage{appendix}

\usepackage{graphicx}% Include figure files
\usepackage{epsfig}

\usepackage{hyperref}
\usepackage{graphicx,epsf}
\usepackage{bm}
\usepackage{amsmath}

\usepackage[dvipsnames]{xcolor}

\newcommand{\be}{\begin{equation}}
\newcommand{\ee}{\end{equation}}
\newcommand{\ba}{\begin{eqnarray}}
\newcommand{\ea}{\end{eqnarray}}
\newcommand{\pard}[2]{\frac{\partial #1}{\partial #2}}
\newcommand{\vc}[1]{{\bm #1}}
\newcommand{\Bgav}[1]{\Big \langle #1 \Big \rangle}
\newcommand{\gav}[1]{\langle #1 \rangle}

\begin{document}

\begin{center}
{\large{\textbf{Nonlinear interaction of Alfv\'enic instabilities and turbulence via the modification of the equilibrium profiles}}}\\
\vspace{0.2 cm}
{\normalsize {A. Biancalani$^{1}$, A. Bottino$^2$, D. Del Sarto$^{3}$, M. V. Falessi$^{4}$, T. Hayward-Schneider$^2$, P. Lauber$^2$, A. Mishchenko$^5$, B. Rettino$^2$, J.N. Sama$^{3}$, F. Vannini$^2$, L. Villard$^6$, X. Wang$^2$, F. Zonca$^{4,7}$\\}}
\vspace{0.2 cm}
\small{
$^1$L\'eonard de Vinci P\^ole Universitaire, Research Center, 92916 Paris La D\'efense, France\\
$^2$Max-Planck Institute for Plasma Physics, 85748 Garching, Germany\\
$^3$Institut Jean Lamour- UMR 7168, University of Lorraine - BP 239, F-54506 Vandoeuvre les Nancy, France\\
$^4$Center for Nonlinear Plasma Science and ENEA C. R. Frascati, 00044 Frascati, Italy\\
$^5$Max-Planck Institute for Plasma Physics, 17491 Greifswald, Germany\\
$^6$Ecole Polytechnique F\'ed\'erale de Lausanne (EPFL), Swiss Plasma Center (SPC), CH-1015 Lausanne, Switzerland\\
$^7$Institute for Fusion Theory and Simulation and Department of Physics, Zhejiang Univ.,
Hangzhou, China
}
\end{center}

\begin{abstract}
Nonlinear simulations of Alfv\'en modes (AM) driven by energetic particles (EP) in the presence of turbulence are performed with the gyrokinetic particle-in-cell code ORB5. The AMs carry a heat flux, and consequently they nonlinearly modify the plasma temperature profiles. The isolated effect of this modification on the dynamics of turbulence is studied, by means of electrostatic simulations. We find that turbulence is reduced when the profiles relaxed by the AM are used, with respect to the simulation where the unperturbed profiles are used. This is an example of indirect interaction of EPs and turbulence. First, an analytic magnetic equilibrium with circular concentric flux surfaces is considered as a simplified example for this study. Then, an application to an experimentally relevant case of ASDEX Upgrade is discussed.
\end{abstract}

%\vskip 6em

\section{Introduction}
\label{sec:intro}
% 
% The GK PIC code ORB5 is used~\cite{Jolliet07,Bottino11}.
% 
% The goal here is to study the effect of the presence of an AM in the dynamics of turbulence.
% We do this in the following way:\\
% First, we perform an electromagnetic simulation like in the EPS-2019 case (results published in the PPCF-2021). For this first simulation, we choose the same geometry as in the EPS-2019 case, but with slightly different profiles (with the new profiles, the BAE is closer to the mid-radius, where the turbulence is peaked). In this first simulation, the AM is observed to modify the equilibrium profiles, due to its heat flux.\\
% Secondly, we take the profiles modified by the AM, and we use them to run some electrostatic ITG-turbulence simulations (without EPs and AMs). We observe that the turbulence level is reduced when we use these modified profiles, with respect to the simulations with the original profiles. This is because the heat flux of the AMs flattens the temperature profiles, therefore reducing the drive for ITG turbulence.

Turbulence develops in tokamak plasmas due to the difference in the values of the equilibrium temperature and density between the core and the edge. Due to the presence of turbulence, the heat fluxes are enhanced, and the confinement is reduced. For this reason, the mitigation of turbulence is considered a key step towards the achievement of controlled nuclear fusion in magnetic confinement devices. Typically, tokamak turbulence generates by the nonlinear interaction of micro-instabilities like ion-temperature-gradient (ITG) modes~\cite{Rudakov65}.
Zonal, i.e. axisymmetric, flows develop in the presence of turbulence, via nonlinear generation, and are one of the main mechanisms of turbulence saturation~\cite{Hasegawa79,Rosenbluth98}.
A population of energetic particles (EP) is also present in tokamak plasmas due fusion reactions and external heating mechanisms. EPs can drive  electromagnetic (EM) oscillations like Alfv\'en Modes (AM)~\cite{Appert82,Chen16} unstable. AMs can redistribute the energetic particles and affect the heating mechanisms aiming at increasing the temperature in the tokamak core.

In the past decades, separate studies have been carried out to study some of these interactions: for example, the transport of energy and particles of the bulk plasma, in the presence of turbulence; and the transport of energy and particles of the EP population, in the presence of AMs. The study of the selfconsistent interaction of EPs, macroscopic AMs and microscopic ITG-turbulence has been for decades a too numerically demanding problem, due to its multi-scale character.
More recently, the need to investigate this interaction has been emphasized by experimental evidences. In particular, a reduction of turbulence in the presence of EPs has been observed for example in AUG, DIII-D, and JET~\cite{Tardini07,Heidbrink09,Romanelli10,Bock17}. Note that the experimental evidence of the interaction of AMs and turbulence (in the absence of EPs) had already been documented in Ref.~\cite{Maraschek97}. 

The construction of theoretical models to investigate the interaction of EPs and turbulence has taken advantage of analytical theory and numerical simulations.
Some milestones in the analytical investigation have been discussed in Ref.~\cite{White89, Chen16, Zonca15, Qiu16}. Flux-tube numerical simulations have also been performed (see for example Ref.~\cite{Angioni09, Bass10, Zhang10, Citrin13, Garcia15, DiSiena19}). Finally, in the last few years, we have also been able to perform global electromagnetic numerical simulations~\cite{BiancalaniEPS19,Biancalani21,DiSiena21,Ishizawa21}.
Different mechanisms can be responsible for the interaction of EPs and turbulence. For example, EPs can directly modify the linear ITG dispersion relation~\cite{DiSiena21}. Or they can drive AMs unstable, which can nonlinearly interact with the ITGs via wave-wave coupling~\cite{Chen22,Qiu23}. Another example is provided by AMs (driven unstable by EPs) exciting ZFs~\cite{Qiu16,Biancalani20}, which can then mitigate turbulence~\cite{BiancalaniEPS19,DiSiena19}. EPs have been also shown to drive the generation of zonal flows via a nonlinear synchronisation process involving Trapped Electron Modes and the low-frequency branch of ITGs, represented by Trapped Ion modes~\cite{Ghizzo22}.
AMs can also indirectly affect turbulence by nonlinearly modifying the equilibrium profiles~\cite{Chen16,Falessi19,ZoncaJPCS21,Biancalani21,Falessi23}. 

In this work, we isolate and investigate this last mechanism in details. This is done by means of the following simplified test. First, we run global selfconsistent electromagnetic simulations of AMs and turbulence (similarly to Ref.~\cite{BiancalaniEPS19,Biancalani21}) and save the profiles modified by the AM. Secondly, we use the modified profiles, for ES simulations of ITG turbulence.
The numerical tool used to perform the numerical simulations is the multispieces EM GK particle-in-cell code ORB5~\cite{Mishchenko19,Lanti20}.
The paper is structured as follows. Sec.~\ref{sec:model} describes the model used for the numerical simulations. In Sec.~\ref{sec:equil-profs}, the equilibrium magnetic field and plasma profiles used for the main study presented here are shown. For continuity with previous work, these are chosen very similar to those of Ref.~\cite{BiancalaniEPS19,Biancalani21}. In Sec.~\ref{sec:EM-sim}, the self-consistent nonlinear electromagnetic simulations are presented, and the nonlinearly modified plasma profiles are measured. In Sec.~\ref{sec:ITG-lin} and Sec.~\ref{sec:ITG-NL}, respectively the linear and nonlinear dynamics of ITG driven by the modified profiles are shown. In Sec.~\ref{sec:NLED-AUG}, an application to a more experimentally relevant case of AUG is shown. Finally, Sec.~\ref{sec:conclusions} is devoted to a summary of conclusions and discussion.

\section{The model}
\label{sec:model}

The numerical tool used here is the girokinetic particle-in-cell code ORB5, originally written for electrostatic turbulence studies~\cite{Jolliet07}, and then extended to its electromagnetic multispecies version~\cite{Bottino11,Mishchenko19,Lanti20}.
ORB5 is global, i.e. it resolves modes with structure comparable with the minor radius.
Thus, it is appropriate for studying AMs with low toroidal
mode number.

In this paper, we use two versions of the code. In the first part of the numerical experiment, we run selfconsistent nonlinear simulations of AMs driven by EPs in the presence of turbulence, therefore the electromagnetic version of the code is used, as in Ref.~\cite{BiancalaniEPS19,Biancalani21}. In the second part of the numerical experiment, we want to study the isolated effect of the nonlinearly modified profiles on ITG turbulence, therefore we use the electrostatic version of the code to avoid the development of AMs.

In the electromagnetic version of the code,
the magnetic potential is split into the Hamiltonian and symplectic parts $A_{\|} = A_{\|}^{{\rm(h)}} + A_{\|}^{{\rm(s)}}$ (see Ref.~\cite{Mishchenko14} for details). The perturbed equations of motion in mixed-variable formulation are~\cite{Mishchenko19}:
\begin{eqnarray}
\label{dotR1}
  &&{} \dot{\vc{R}}^{(1)} = \frac{\vc{b}}{B_{\|}^*} \times \nabla \Bgav{ \phi -
    v_{\|} A_{\|}^{\rm(s)} - v_{\|} A_{\|}^{\rm(h)} } -
  \frac{q}{m} \,\gav{A^{\rm(h)}_{\|}} \, \vc{b}^* \\
  \label{dotp1}
  &&{} \dot{v}_{\|}^{(1)} = \,-\, \frac{q}{m} \,
  \left[ \vc{b}^* \cdot \nabla \Bgav{\phi - v_{\|} A_{\|}^{\rm(h)}} + \pard{}{t}
    \Bgav{A_{\|}^{\rm(s)}} \right]
   -  \frac{\mu}{m} \, \frac{\vc{b} \times \nabla B}{B_{\|}^*} \cdot \nabla
    \Bgav{A_{\|}^{\rm(s)}}
 \end{eqnarray}
where
\begin{eqnarray} 
 \label{bstar}
  &&{}\vc{b}^* = \vc{b}^*_0 + \frac{\nabla\gav{A_{\|}^{{\rm(s)}}}\times \vc{b}}{B_{\|}^*} \ , \;\;\; \vc{b}^*_0 = \vc{b} + \frac{m v_{\|}}{q B_{\|}^*} \nabla\times\vc{b} \\
  &&{} B_{\|}^* = B + \frac{m v_{\|}}{q} \vc{b}\cdot\nabla\times\vc{b}
\end{eqnarray}
We also have an equation for $\partial A_{\|}^{\rm(s)} / \partial t$ (Ohm's law, see \cite{Mishchenko14}):
  \be
  \label{Ohm}
  \pard{}{t}A_{\|}^{\rm(s)} + \vc{b} \cdot \nabla \phi = 0
  \ee
  %
  %The zeroth-order gyrocenter characteristics remain unchanged.
  %The mixed-variable distribution function is solved from the gyrokinetic Vlasov
  %equation.
  %
  and the field equations (here we use the notation as in Ref.~\cite{Mishchenko14}):
  \ba
  \label{amp}
  &&{}
  \label{eq:GKPoisson}
  \,-\,\nabla\cdot\left(\frac{n_0}{B \omega_{ci}}\nabla_{\perp}\phi\right)
   = \bar{n}_{1i} - \bar{n}_{1e} \\
  \label{eq:Ampere}
  &&{}
  \sum_{s = i,e}\frac{\beta_s}{\rho_{s}^{2}} A_{\|}^{\rm(h)}
  - \nabla_{\perp}^{2} A_{\|}^{\rm(h)}
  = \mu_{0} \sum_{s = i,e} \bar{j}_{\|1s} + \nabla_{\perp}^2 A_{\|}^{\rm(s)}
  \ea
  
For the electrostatic simulations, only the thermal ions and energetic ions (the EP species) are treated kinetically, whereas the electrons are treated adiabatically. For these simulations, only the scalar potential is needed, so the gyrokinetic Poisson law, Eq.~\ref{eq:GKPoisson}, is solved.

For noise control purposes, as well as for maintaining some of the plasma profiles close to their initial state, a modified Krook operator is used for the thermal ions and electrons (not for the EPs):
\begin{equation}
\frac{\mathrm{d} f_s}{\mathrm{d} t} = S(f_s)
\end{equation}
with $S(f_s)=-\gamma_K(f_s-f_{0,s}) + S_\mathrm{corr}(f_s)$.
The coefficient $\gamma_K$ is chosen empirically such that the signal to noise ratio is maintained at a sufficiently high level, while only weakly affecting the physics of interest. It is typically chosen as 5\% or 10\% of the maximal growth rate. Here, we use the same value as in Ref.~\cite{Biancalani21}, because the dynamics is very similar. The term $S_\mathrm{corr}(f_s)$ is such that a number of moments $M(v)$ are conserved by the source term:
$\langle\int S(f_s)M(v)d^3v\rangle=0$, where $\langle Q \rangle$ stands for the flux-surface average of a quantity Q. $S_\mathrm{corr}$ is used to conserve the undamped ExB Zonal Flow residual. For more details see Refs.~\cite{McMillan08,Lanti20}.

% 
% The goal here is to study the effect of the presence of an AM in the dynamics of turbulence.
% We do this in the following way:\\
% First, we perform an electromagnetic simulation like in the EPS-2019 case (results published in the PPCF-2021). For this first simulation, we choose the same geometry as in the EPS-2019 case, but with slightly different profiles (with the new profiles, the BAE is closer to the mid-radius, where the turbulence is peaked). In this first simulation, the AM is observed to modify the equilibrium profiles, due to its heat flux.\\
% Secondly, we take the profiles modified by the AM, and we use them to run some electrostatic ITG-turbulence simulations (without EPs and AMs). We observe that the turbulence level is reduced when we use these modified profiles, with respect to the simulations with the original profiles. This is because the heat flux of the AMs flattens the temperature profiles, therefore reducing the drive for ITG turbulence.

\newpage
\section{Magnetic equilibrium and plasma profiles}
\label{sec:equil-profs}

For continuity with the previous work, we choose the magnetic equilibrium and plasma profiles very similar to those used in the case labelled here as the ``EPS-2019 case'', published in Ref.~\cite{BiancalaniEPS19, Biancalani20, Biancalani21}. The two only differences are the shape of the q-profile and the localization of the density and temperature gradients, as described below.

\begin{figure}[b!]
\begin{center}
\includegraphics[width=0.55\textwidth]{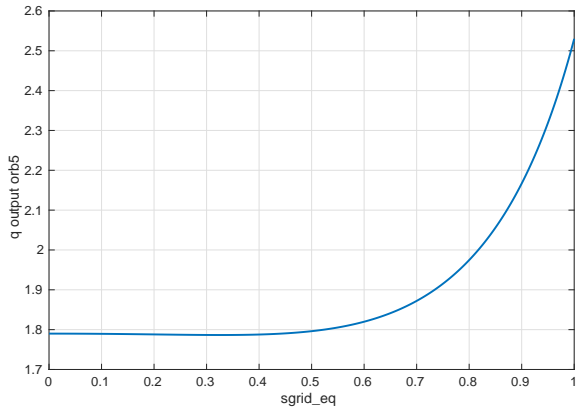}
\vskip -1em
\caption{\label{fig:q_s} Safety factor profile in s coordinate.}
\end{center}
\end{figure}

Like in the EPS-2019 case, a magnetic equilibrium with inverse aspect ratio $\epsilon=0.1$ is considered (the major radius is $R_0 = 10$ m, the minor radius is $a=1.0$ m), with circular concentric flux surfaces. The magnetic field on axis is $B_0 = 3.0$ T.
Differently from the EPS-2019 case, the q profile is nearly monotonic here (see Fig.~\ref{fig:q_s}), with a rational surface at mid-radius, allowing a better localization of the modes of interest. In particular, we have a value of $q(0)$=1.79 at the axis, a minimum of $q_{min}=1.787$ at $s=0.33$, a value of $q=1.8$ at $s$ = 0.525, and a value at the edge of $q(1)$ = 2.53. Here, the flux radial coordinate $s$ is defined as $s=\sqrt{\psi_{pol}/\psi_{pol}(edge)}$. The rational surface at s=0.525 corresponds to a normalized radius chosen as reference position, with value $\rho_r=0.5$. Here $\rho$ is a normalized radial coordinate defined as $\rho=r/a$.

In tokamaks, the type of turbulence under investigation strongly depends on the dimensionless parameter $\rho^* = \rho_s/a $ (with $\rho_s = \sqrt{T_e/m_i}/\Omega_i$ being the sound Larmor radius).
Here, like in the EPS-2019 case, we choose a value of $\rho^*$ similar to the CYCLONE base case (originally chosen as an international benchmark case for ITG turbulence in a DIII-D configuration): $\rho^* = \rho_s/a = 0.00571$ (therefore $Lx=2/\rho^* = 350$). For comparison, note that $\rho^* = 1/100$ in Ref.~\cite{Ishizawa21}.
The electron thermal to magnetic pressure ratio of $\beta_e = 8\pi \langle n_e \rangle T_e(\rho_r)/B_0^2 = 5\cdot 10^{-4}$ (with $\langle n_e \rangle$ being the volume averaged electron density).
%Here $\langle n_e \rangle = $ \verb|nbar|$_e * n_e(\rho_r)$, with \verb|nbar|$_e$ = 0.971 given in ORB5's standard output (we have \verb|nsel_profile=2| in the input file, therefore \verb|nbar| is calculated at the reference position \verb|speak| and not on the axis), and depends on the choice of $\kappa_n$ (see Eq.~\ref{eq:n_EP}) relative to the electron density profile.
The equilibrium density and temperature profiles are different in the two parts of this study, and they are defined in the following way.

1) In the first part, we use very similar equilibrium density and temperature profiles as in the EPS-2019 case. The initial profiles are described by this equation:
\begin{equation}\label{eq:n_EP}
n(\rho)/n(\rho_r) = \exp [-\Delta \; \kappa \; \tanh ((\rho-\rho_r)/\Delta)] 
\end{equation}
where both thermal ions and electrons have $\kappa_n = 0.3$ and $\kappa_t = 1.0$ respectively for the density and the temperature profiles.
The only difference here, with respect to the EPS-2019 case, is that the density and temperature gradients are slightly more localized around mid-radius. This is done by selecting, for both thermal ions and electrons, a slightly smaller value of $\Delta$: $\Delta = 0.15$.
In this first part of the study, we run nonlinear electromagnetic simulations of turbulence, zonal flows, AMs and EPs, similarly to the simulations shown in Ref.~\cite{BiancalaniEPS19,Biancalani21} for the EPS-2019 case.

2) In the second part of this study, the initial density and temperature profiles are not given by Eq.~\ref{eq:n_EP}. On the other hand, we take the temperature and density profiles given in output from ORB5 in the electromagnetic simulations performed in the first part. These profiles are initialized in electrostatic ITG-turbulence simulations (without AMs and EPs).

\section{First part of the study: EM simulation}
\label{sec:EM-sim}

\begin{figure}[b!]
\begin{center}
\includegraphics[width=0.55\textwidth]{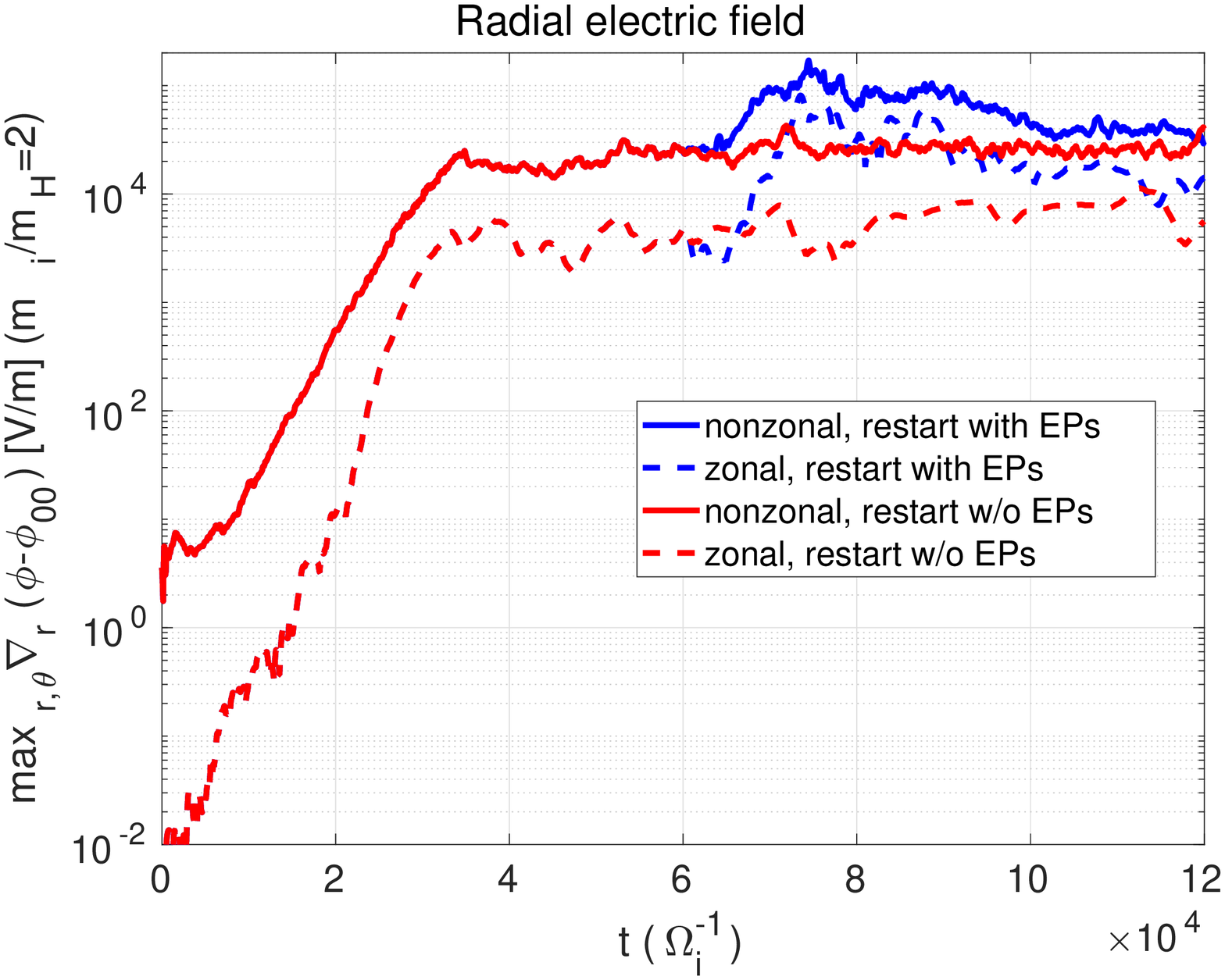}\includegraphics[width=0.5\textwidth]{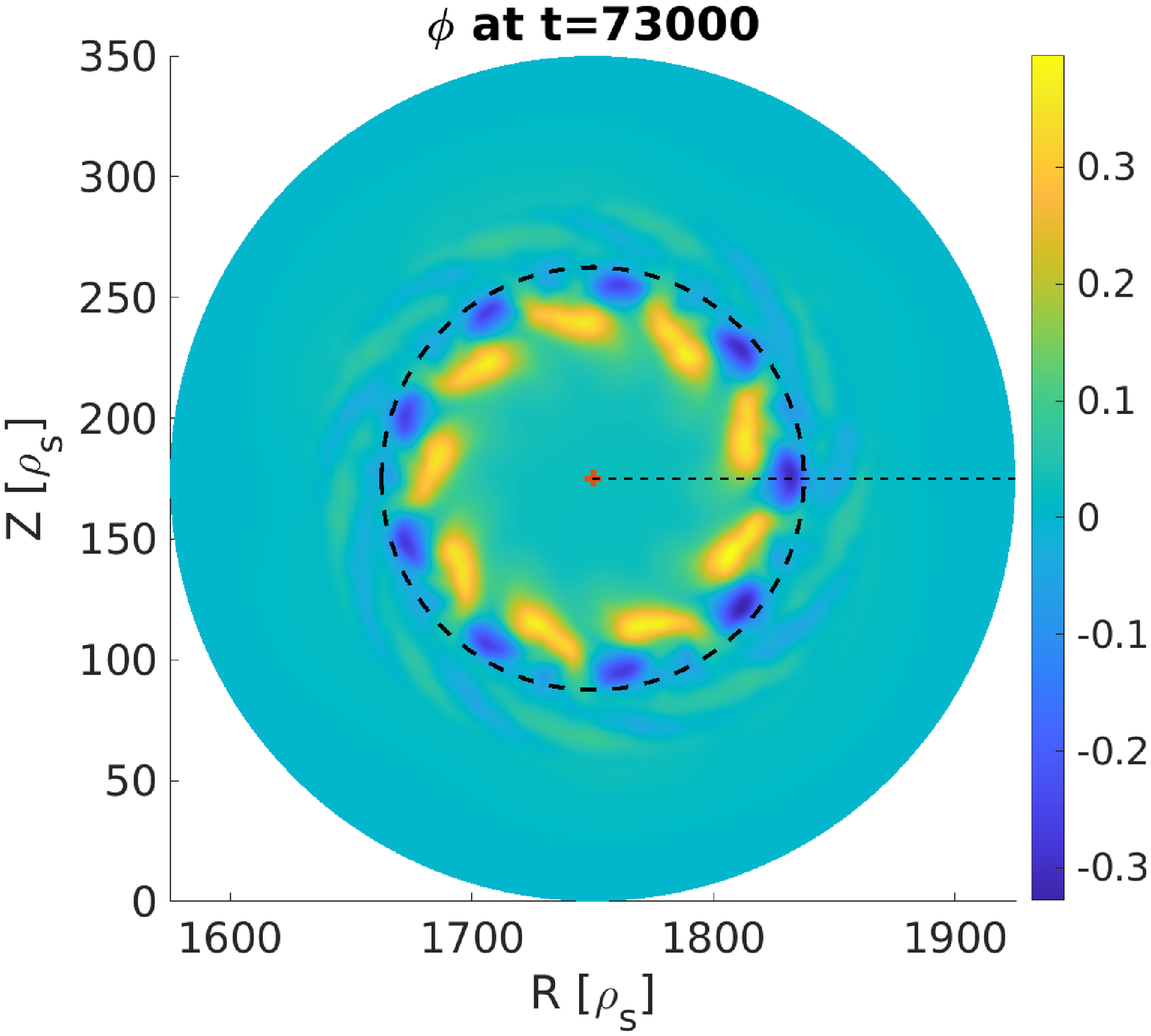}
%\includegraphics[width=0.49\textwidth]{q1_4-gamma_e-k2.eps}
%\vskip -1em
\caption{\label{fig:monotq-EMsim-Er_t} Left: evolution of the radial electric field in time for a simulation with EPs (blue) and without EPs (red). In the simulation with EPs, an AM is driven unstable. Right: structure of the AM (left) and ZF (right). The AM is identified as a BAE.}
\end{center} 
\end{figure}

In this section, we show the result of the nonlinear selfconsistent EM simulation. The dynamics is very similar to that shown in Ref.~\cite{BiancalaniEPS19,Biancalani21}. The slightly different profiles are found to affect the radial localization of the AM. The evolution of the fields in time can be observed in Fig.~\ref{fig:monotq-EMsim-Er_t}-left. EPs are switched on at $t=6\cdot 10^4 \, \Omega_i^{-1}$.
The AM is a beta-induced Alfv\'en Eigenmode (BAE) with n=5, m=9 (see Fig.~\ref{fig:monotq-EMsim-Er_t}-right).
%The BAE excites a zonal flow due to forced-driven excitation (see Fig.~\ref{fig:monotq-EMsim-structure}-right)

% 
% \begin{figure}[h!]
% \begin{center}
% \includegraphics[width=0.48\textwidth]{FIGS/monotq-EMsim-phi_RZ-t-73000.eps}
% \includegraphics[width=0.48\textwidth]{FIGS/monotq-EMsim-zonalEr_s.eps}
% %\includegraphics[width=0.49\textwidth]{q1_4-gamma_e-k2.eps}
% %\vskip -1em
% \caption{\label{fig:monotq-EMsim-structure} Structure of the AM (left) and ZF (right). The AM is identified as a BAE, like in the case of the EPS-2019.}
% \end{center} 
% \end{figure}

\begin{figure}[t!]
\begin{center}
\includegraphics[width=0.48\textwidth]{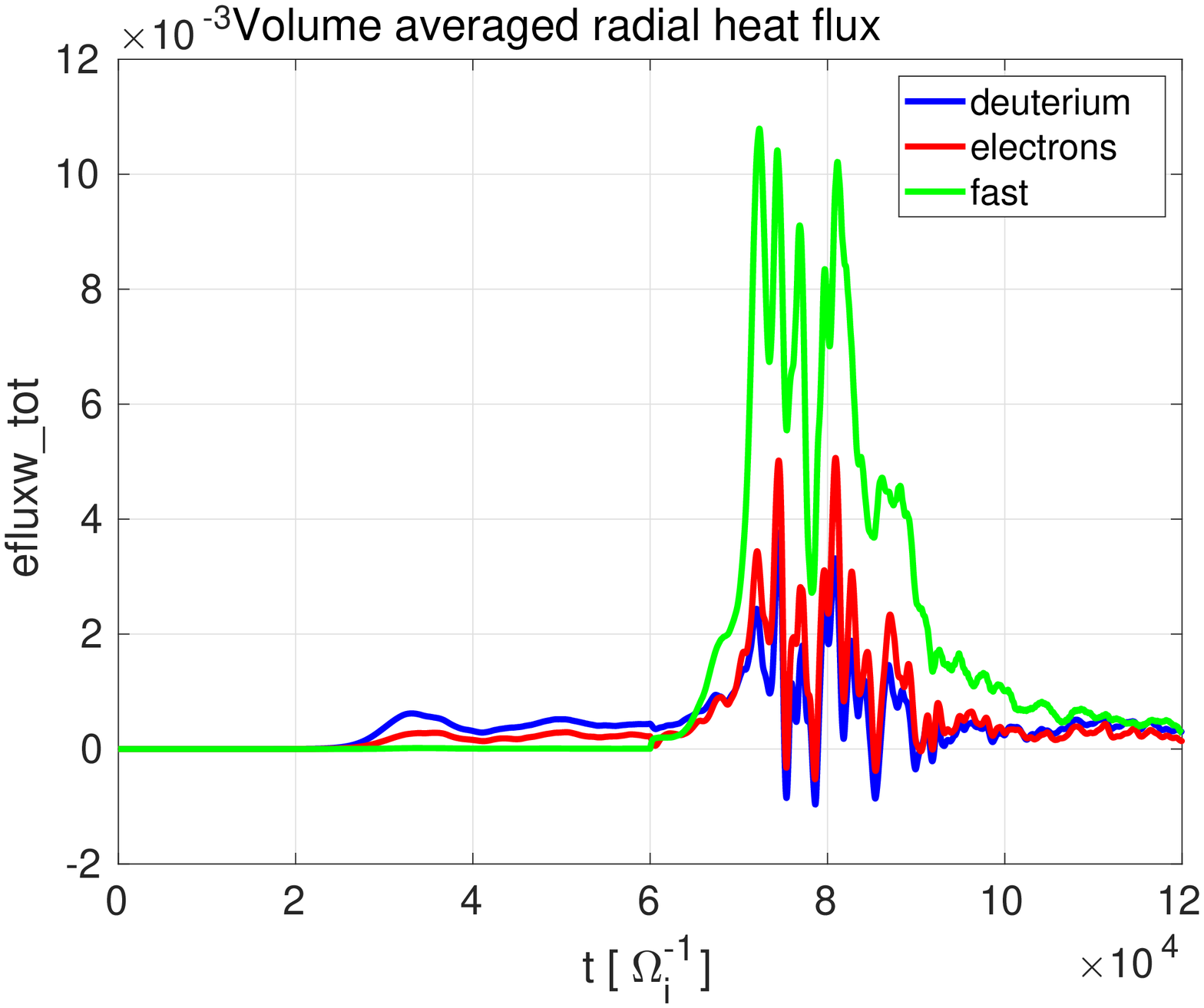}
\includegraphics[width=0.48\textwidth]{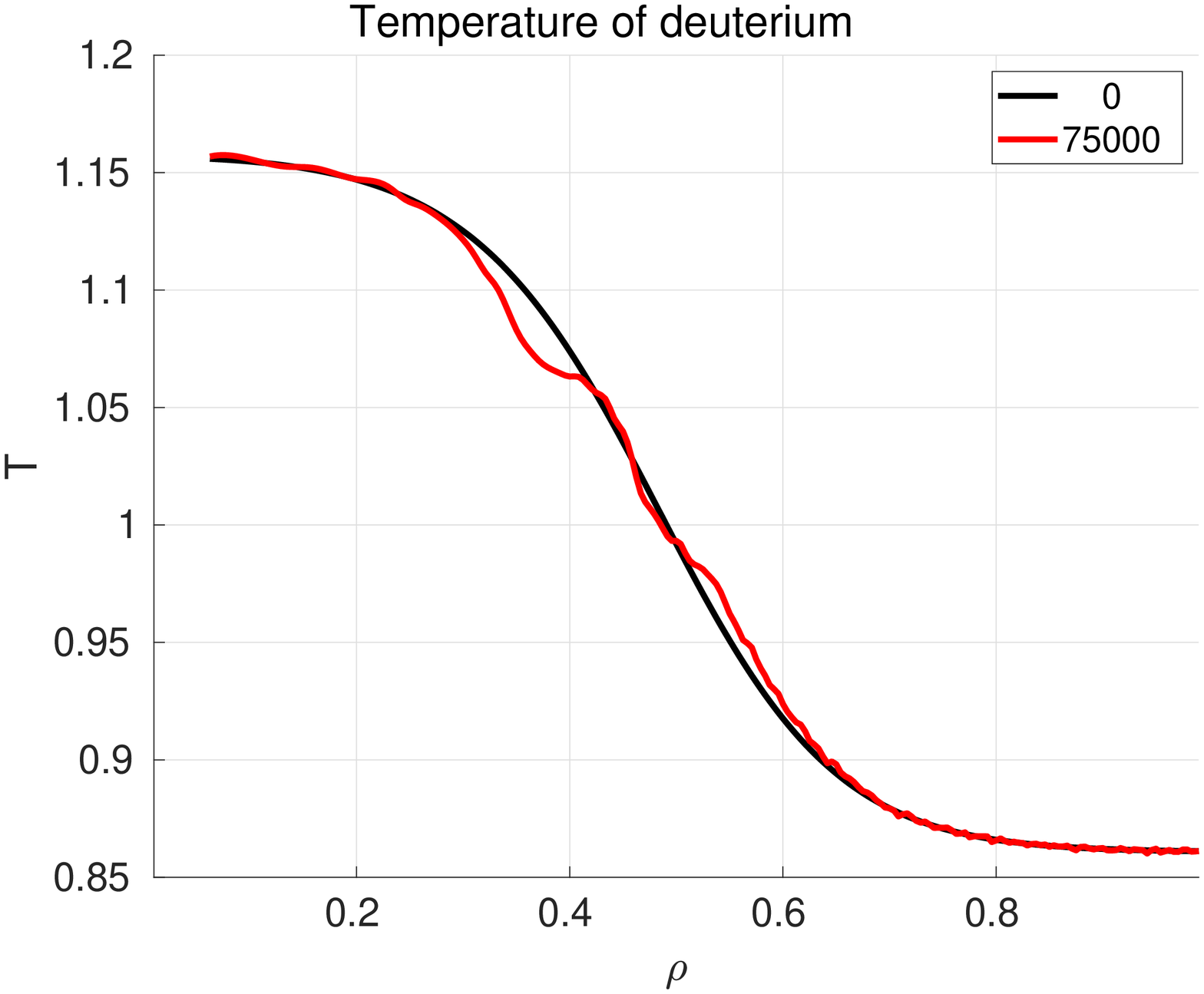}
\vskip -1em
\caption{\label{fig:monotq-EMsim-heatflux} Evolution of the heat fluxes in time (left) for a simulation with EPs. On the right, the result of the effect of the heat fluxes on the modification of the temperature profiles of deuterium (similar effects are found on the electrons). The same profile measured at t=75000 $\Omega_i^{-1}$ is used for the electrostatic simulations in the second part of this study.}
\end{center} 
\end{figure}

Like in the case of Ref.~\cite{BiancalaniEPS19,Biancalani21}, the BAE carries heat fluxes. This can be seen in Fig.~\ref{fig:monotq-EMsim-heatflux}-left. These heat fluxes modify the equilibrium profiles. As an example, the temperature profile of the thermal ions is shown at t=0 and t=75000 $\Omega_i^{-1}$ in Fig.~\ref{fig:monotq-EMsim-heatflux}-right. Note that the temperature is flattened at the radial position of the BAE, i.e. around $\rho=0.4$. Note also that, like shown in Ref.~\cite{Biancalani21}, the direct interaction of EPs and turbulence is negligible in this regime. For completeness, note that the EP transport due to turbulence has been proved to be consistent with  quasilinear theory, and to become negligible for large values of the EP temperature~\cite{White89,Zhang10,Chen16}.

We now take the temperature and density profiles measured at the end of this nonlinear simulation, and we use them as starting conditions for the electrostatic turbulence simulations, discussed in the next section.

\newpage
\section{Second part of the study: A) linear dynamics of ITG modes}
\label{sec:ITG-lin}

\begin{figure}[b!]
\begin{center}
\includegraphics[width=0.48\textwidth]{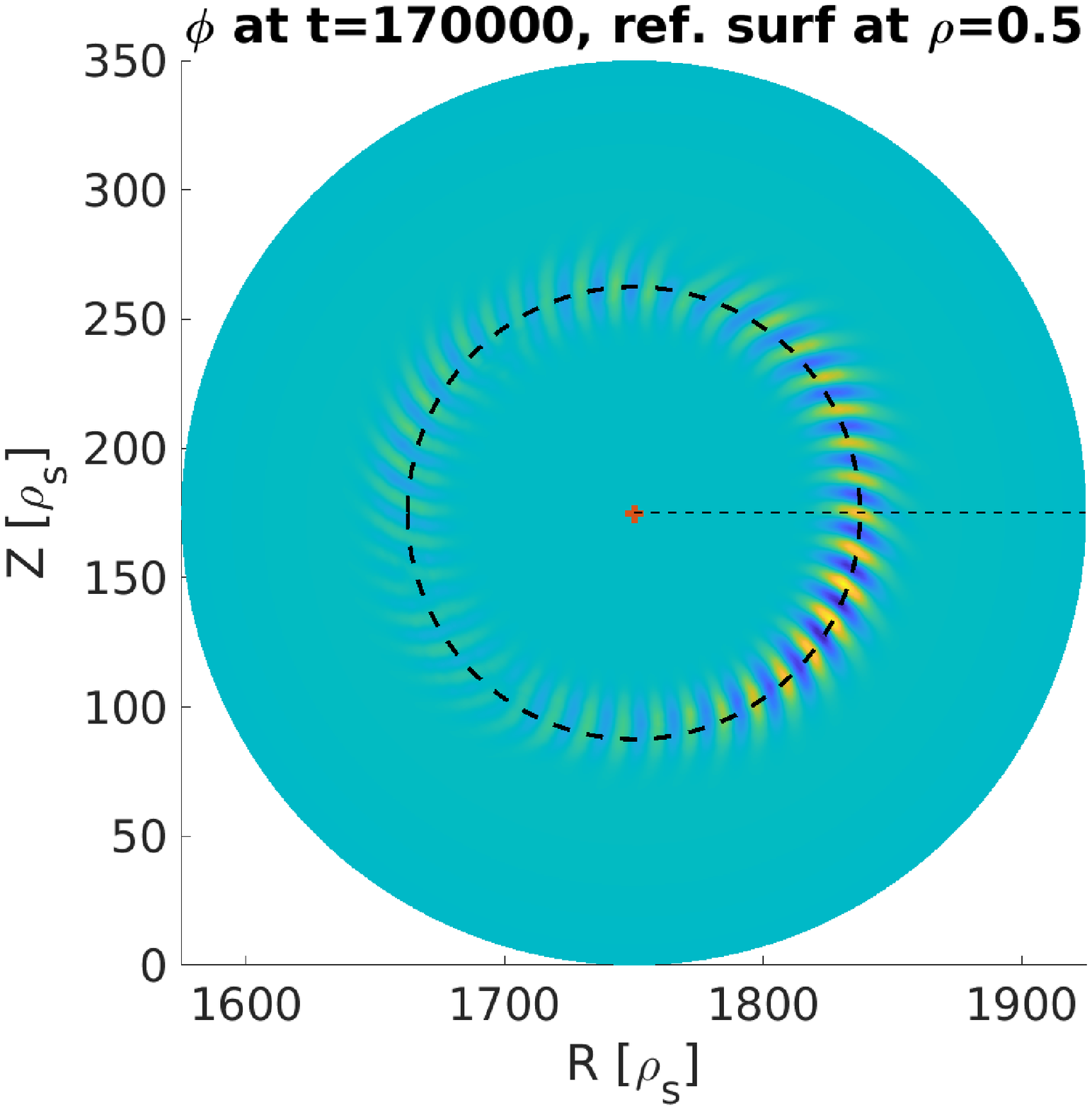}
\includegraphics[width=0.48\textwidth]{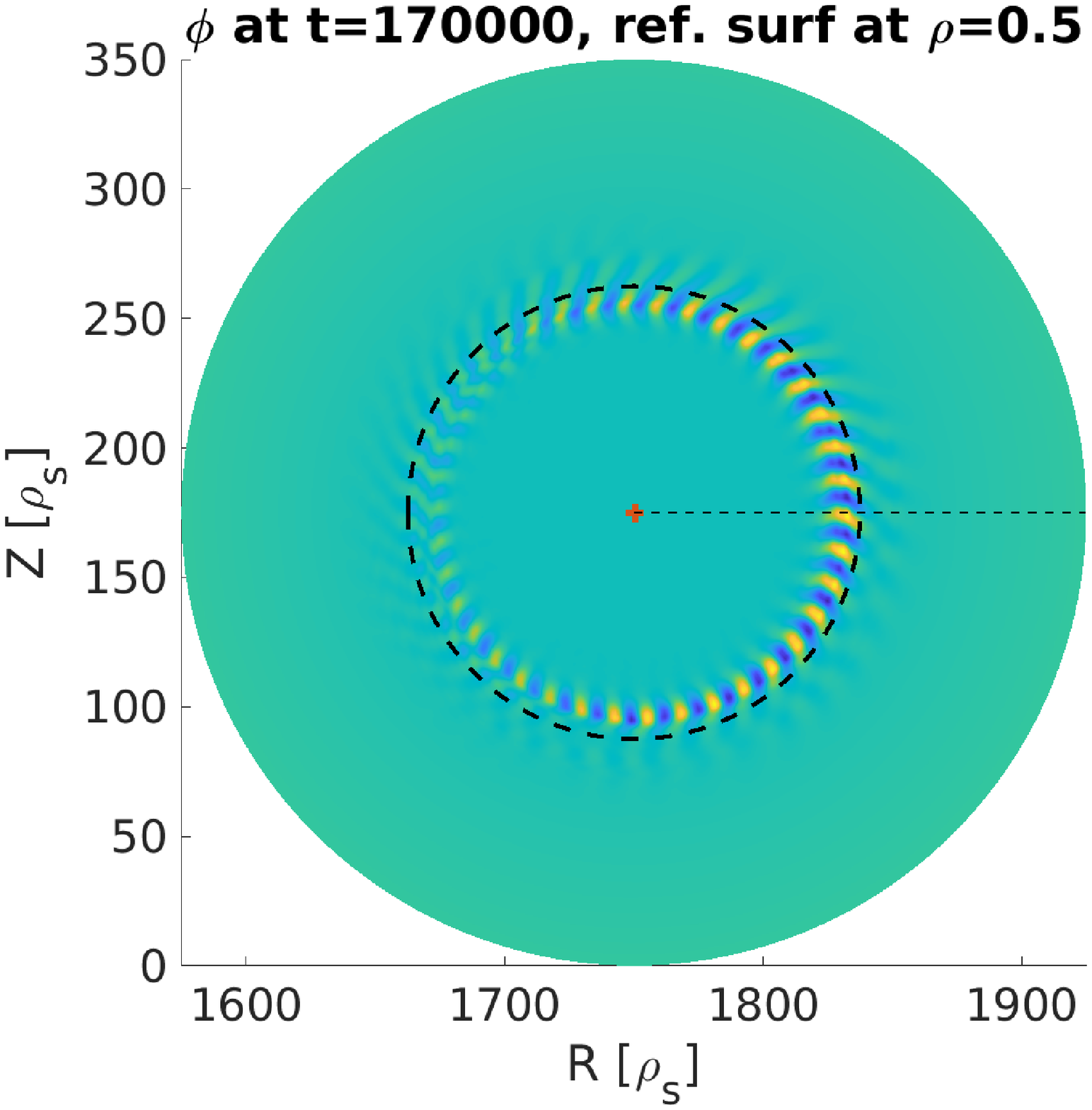}
%\includegraphics[width=0.49\textwidth]{q1_4-gamma_e-k2.eps}
%\vskip -1em
\caption{\label{fig:monotq-ESlin-structure} Structure of the ITG dominant mode in linear simulations with original unperturbed profiles (left) and with the profiles modified by the AM (right).}
\end{center} 
\end{figure}

In this section, we show the results of linear electrostatic simulations (therefore without AMs). We also keep only two species in the dynamics: thermal ion and electrons. No EPs are initialized. Two cases are compared: in one case, we load the original profiles (i.e. as in Sec.~\ref{sec:EM-sim}); in the other case, we take the profiles of the EM simulation, modified by the presence of the AM, and we use these for the ES turbulence simulation. The goal is to see how the dynamics of the ITG modes differs in the two cases.

These linear simulations show that both the structure and the growth rate of the ITGs is different in the two kinds of profiles. The structure is shown in Fig.~\ref{fig:monotq-ESlin-structure}. The maximum linear growth rate for the case with the original (i.e unperturbed) profiles is measured as $\gamma_{lin}\simeq 1.2\cdot 10^{-4} \Omega_i$. The maximum linear growth rate for the case with the profiles modified by the AM mode is measured as $\gamma_{lin}\simeq 1.0\cdot 10^{-4} \Omega_i$. Therefore, ITGs are  found to be linearly mitigated by the presence of the AM.

\section{Second part of the study: B) Nonlinear dynamics of ITG turbulence}
\label{sec:ITG-NL}

\begin{figure}[h!]
\begin{center}
\includegraphics[width=0.52\textwidth]{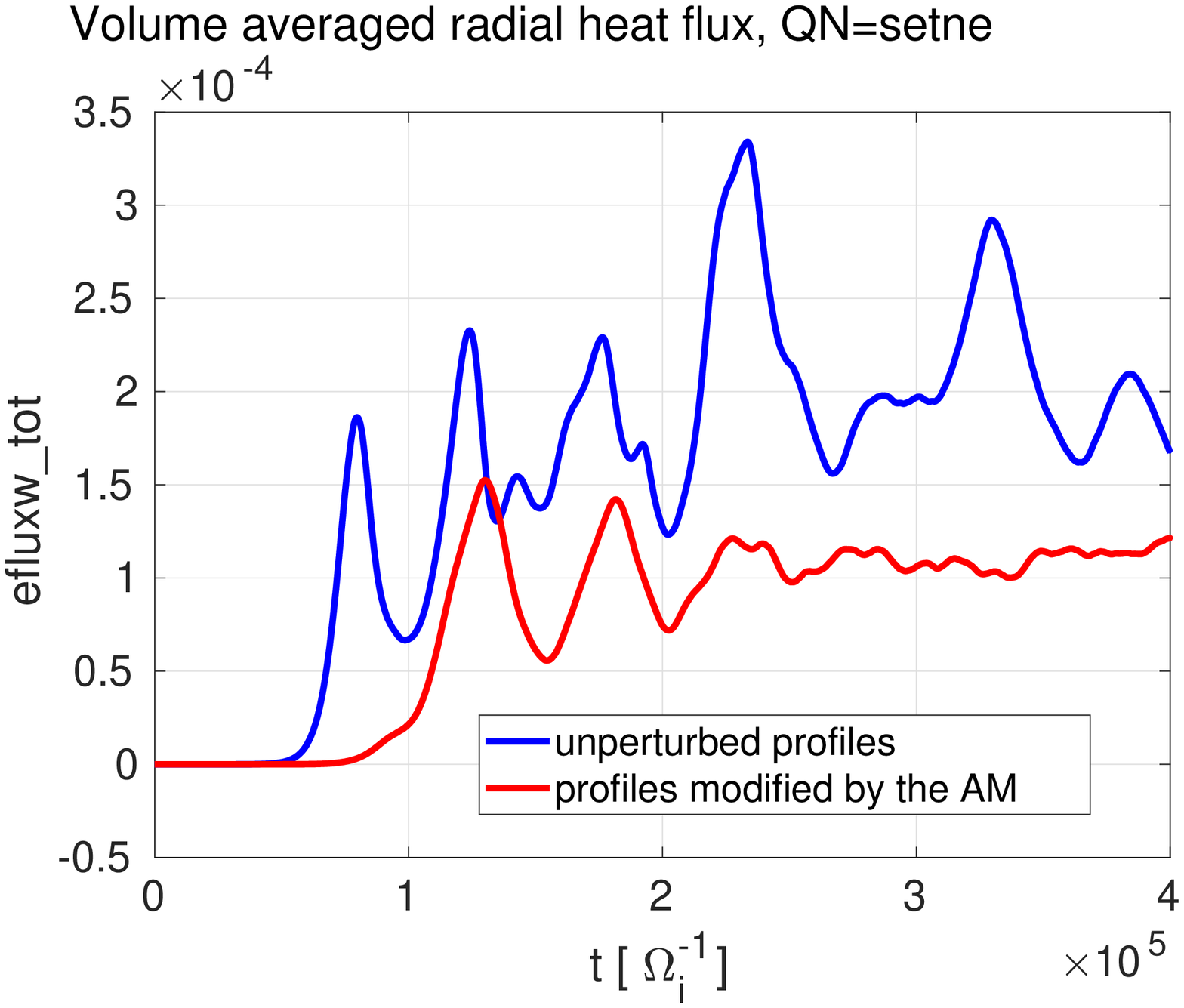}\includegraphics[width=0.5\textwidth]{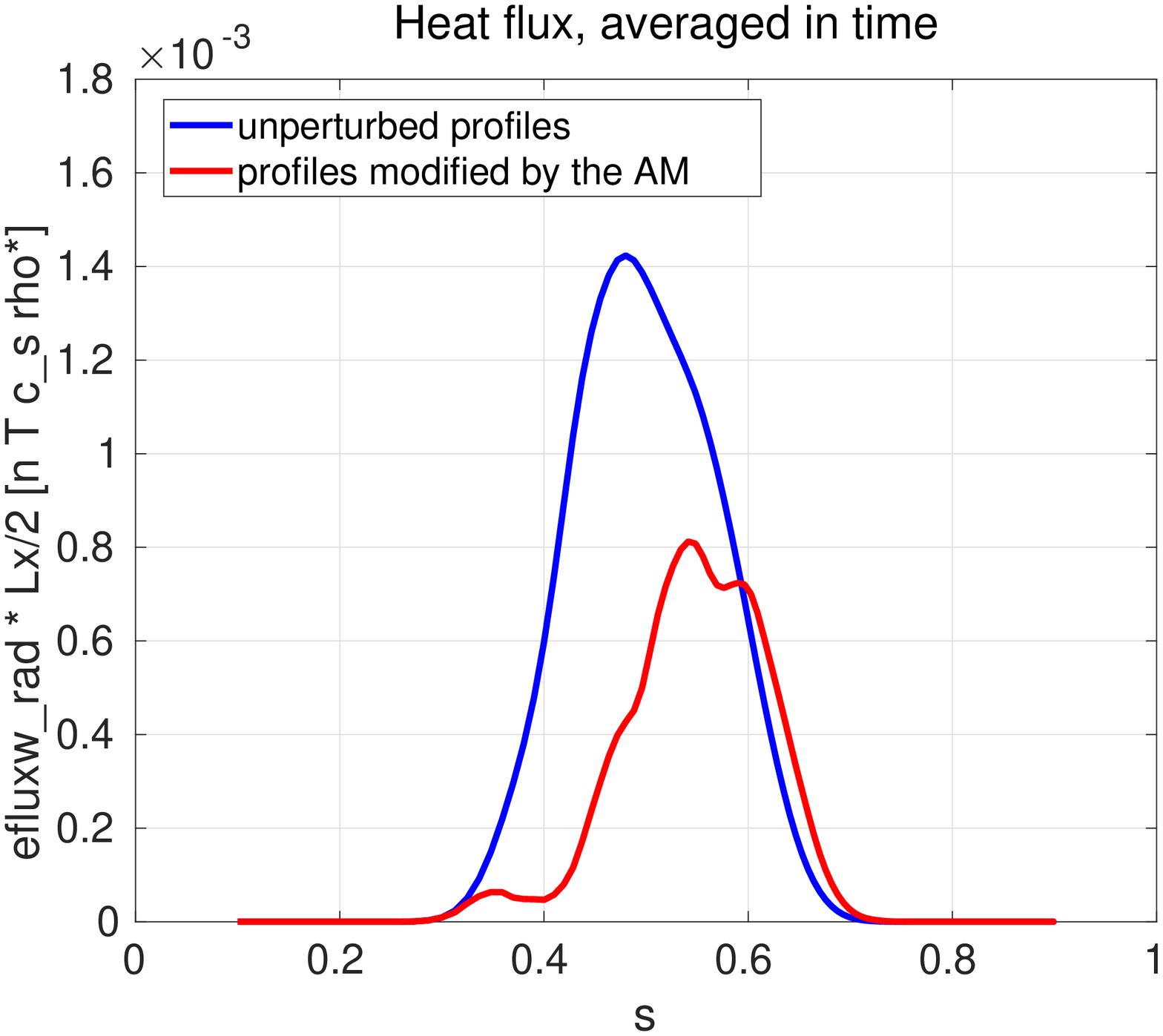}
%\includegraphics[width=0.49\textwidth]{q1_4-gamma_e-k2.eps}
%\vskip -1em
\caption{\label{fig:monotq-ESNL-efluxw} Turbulence intensity measured with the ion heat fluxes.  The two cases with original (blue) and with the modified (red) profiles are depicted. On the left, one can see the time evolution of the heat flux. Note that the heat flux is lower when the modified profiles are used. On the right, the radial structure of the heat flux, averaged in time.}
\end{center} 
\end{figure}

In this section, we show the results of nonlinear electrostatic simulations. Exactly the same case as shown in Sec.~\ref{sec:ITG-lin} is considered. We want to compare the dynamics with unperturbed profiles, and with profiles modified by the heat flux carried by the AM. A possible way to give an estimation of the turbulence intensity is by measuring the heat flux. In Fig.~\ref{fig:monotq-ESNL-efluxw}-left, the time evolution of the ion heat flux of the nonlinear  ITG simulations is shown for both cases. Note that the heat flux in the simulation with modified profiles is about a factor 2 lower than the heat flux of the simulation with unperturbed profiles.
In Fig.~\ref{fig:monotq-ESNL-efluxw}-right, we can see that the  time averaged radial heat flux is nearly suppressed at the radial position of the AM, where the profiles are flattened. 
In summary, we can observe an indirect mechanism of turbulence reduction of the EPs (by means of the AM).

\newpage
\section{Application to an experimental case: the effect of AMs on the equilibrium profiles in the NLED-AUG case}
\label{sec:NLED-AUG}

In the previous sections, we have selected a tokamak configuration and we have performed selfconsistent electromagnetic simulations, measured the nonlinearly modified temperature profile, and used these profiles to study the effect on ITG linear and nonlinear dynamics. For continuity with previous work, the chosen case is a tokamak configuration with low inverse aspect ratio, circular concentric flux surfaces, and relatively low beta. This case allows selfconistent electromagnetic simulation including multiple scales, at a relatively low numerical cost.

In this section, we want to draw some conclusions on more experimentally relevant scenarios. We take the NLED-AUG case~\cite{NLED-AUG} as an example, because the linear and nonlinear physics of AMs has been studied in detail for this case with ORB5~\cite{Vannini20,Vannini21,Vannini22, Rettino22} and benchmarked with other codes~\cite{Vlad21}. Therefore, we can claim that our simulations correctly include the main nonlinear dynamics of these AMs. This preparatory phase is crucial to be able to make statements on the physics of Alfv\'en modes in experimentally relevant predictions.
%We show here that, even in this more experimentally relevant regime, the AMs are capable of carrying a strong heat flux and modify the temperature profiles.

The NLED-AUG simulations presented here use an experimental (shaped) magnetic equilibrium, and experimental density and temperature profiles of all the species. The equilibrium distribution function of the thermal species is a Maxwellian, and that of the EPs is an isotropic slowing-down.
The reader should refer to Ref.~\cite{Vannini22JPCS} for more details on this case. 
In this simulation, modes among n=0 and n=6 are kept. In the linear phase, the most unstable mode is the mode with n=2. 
The simulation is fully nonlinear, meaning that the markers of all species are pushed along their full (equilibrium + perturbation) trajectories.

\begin{figure}[b!]
\begin{center}
\includegraphics[width=0.55\textwidth]{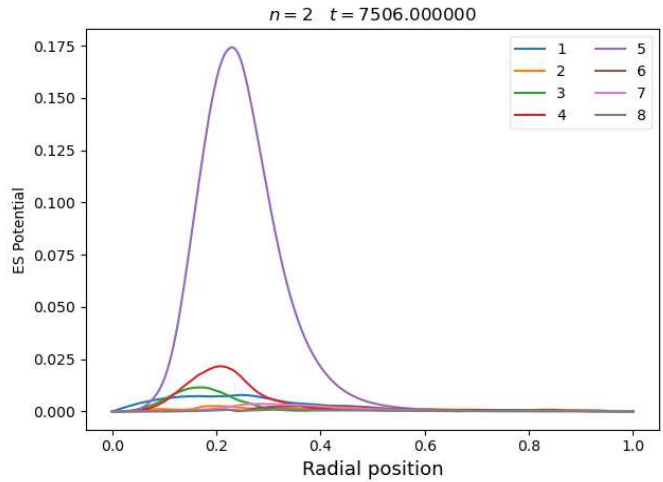}
%\includegraphics[width=0.49\textwidth]{q1_4-gamma_e-k2.eps}
%\vskip -1em
\caption{\label{fig:NLEDAUG-fields} Radial structure of the AM in the NLED-AUG case}
\end{center} 
\end{figure}

\begin{figure}[h!]
\begin{center}
\includegraphics[width=0.49\textwidth]{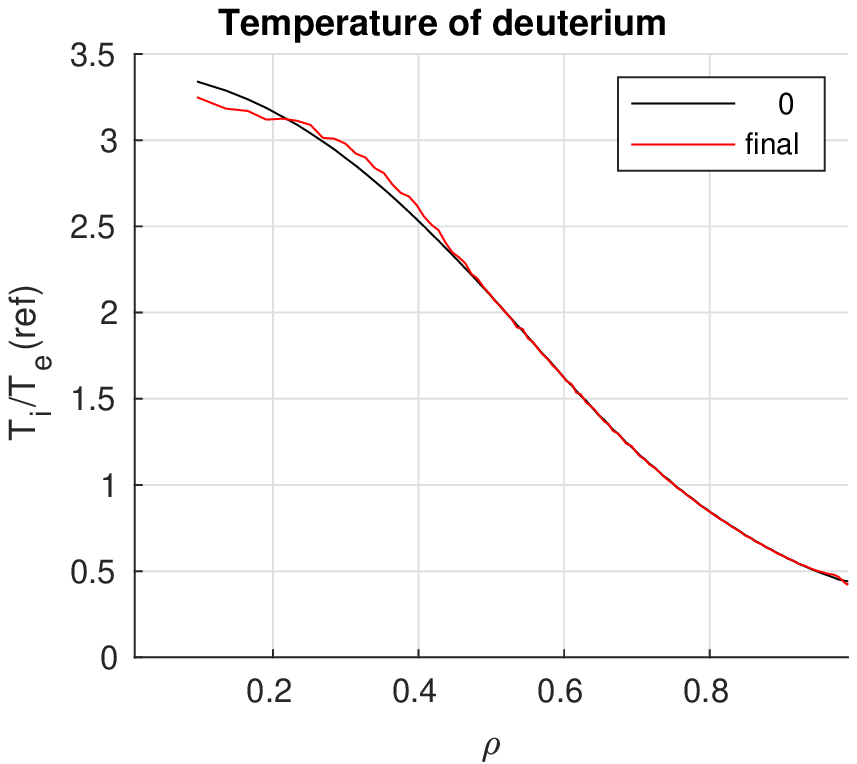}\includegraphics[width=0.49\textwidth]{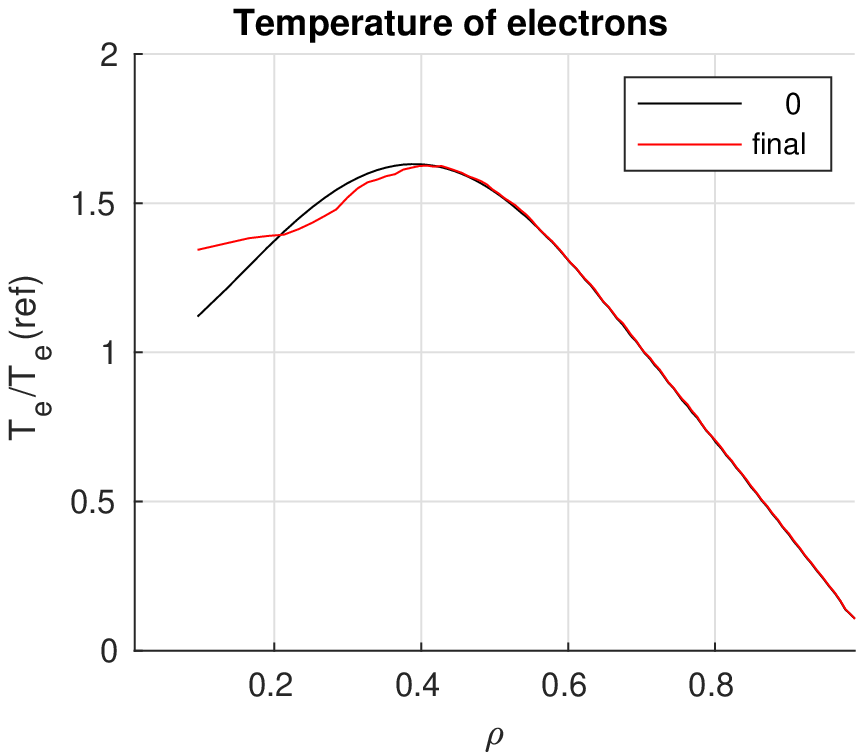}
%\includegraphics[width=0.49\textwidth]{q1_4-gamma_e-k2.eps}
%\vskip -1em
\caption{\label{fig:NLEDAUG-profs} Radial profiles of the temperature of thermal ions (left) and electrons (right) at the initial state of the nonlinear simulations of AMs, and after the nonlinear saturation.}
\end{center} 
\end{figure}

The radial structure of the mode, as given in Refs.~\cite{Vannini22JPCS}, shows a mode peaked in the core, at $s=\sqrt{\psi/\psi_{edge}}~\simeq 0.25$.
%The saturated electric field of this AM in our simulations is:
%\begin{equation}
%\delta E_{r,max}= ???
%\end{equation}
The corresponding modification of the temperature profiles of thermal ions and electrons can be studied. In Fig.~\ref{fig:NLEDAUG-profs}, the temperature profiles normalized with $T_e($ref$)$, at the beginning and at the end of the simulations (with $T_e($ref$)$ being the electron temperature measured at the axis, $\rho=0$, at the beginning of the simulation, $t=0$). Note that a sensible flattening of the temperature profiles is caused by the AM for both ions and electrons. This is especially visible near the AM radial localization, i.e. $s \simeq 0.25$.
As a consequence, we indicate that the mechanism of indirect interaction of EPs and ITG turbulence shown in this paper, can be important not only in simplified configurations, but also in experimentally relevant regimes. We leave the analysis of this AUG shot and the comparison with experimental measurements to a dedicated work.

%\newpage
\section{Conclusions and discussion}
\label{sec:conclusions}

% Here, we have investigated the indirect interaction of EPs and turbulence by means of the heat fluxes carried by the AMs. These heat fluxes modify the equilibrium profiles, and in particular they flatten the profiles in some regions which can be important for the drive of the ITG modes. As a consequence, we measure a reduction of the ITG growht, both linearly and nonlinearly.\\
% 
% \noindent
% \emph{to be completed}

The transport of energy and particle in tokamak plasmas is intrinsically a multi-scale problem, due to the nonlinear interaction of waves and instabilities having different space and time scales. Global modes like AMs, driven unstable by EPs, can nonlinearly excite zonal (i.e. axisymmetric) structures, and indirectly affect the dynamics of ITGs. An example of zonal structures is the zonal radial electric field, linked to zonal poloidal flows. Another example is the perturbation of the equilibrium density and temperature (zonal) profiles.

In this paper, we have investigated this latter possible mechanism of interaction of EPs and ITG turbulence, namely we have investigated how an AM can modify the profiles, and indirectly affect ITG turbulence. We have considered two tokamak cases: the EPS-2019 case (as in Ref.~\cite{BiancalaniEPS19,Biancalani21}), which allows a detailed investigation due to its relatively low computational cost, and the more experimentally relevant NLED-AUG case (as in Ref.~\cite{Vannini22}).

In the EPS-2019 case, we have considered the selfconsistent EM simulation where AMs driven by EPs coexist with zonal structures, and ITG turbulence. We have measured the profiles nonlinearly modified by the AMs, and we have used those to run ES simulations of ITGs. Linear and nonlinear simulations both show an effect of the modified profile: the ITG is mitigated in the modified profile, due to the lower gradient of temperature (which has been flattened by the heat flux carried by the AM).
In the NLED-AUG case, we have shown that AMs can still carry a substantial heat flux, and sufficiently large to flatten the temperature profiles at the location of the AM.
This result paves the way for a different point of view on the interpretation of experimental results such as the turbulence reduction in the presence of EPs, in experimentally relevant cases.

As next steps, selfconsistent EM simulations will be performed in experimentally relevant configurations like the NLED-AUG case, in the direction shown by the recent works like those of Ref.~\cite{Mishchenko21,Mishchenko22,Mishchenko23}.

\section*{Acknowledgments}
%\noindent
%Acknowledgments\\
Interesting discussions with L. Chen, Z. Qiu, E. Poli, B. McMillan, E. Lanti, and N. Ohana are gratefully acknowledged.
This work has been carried out within the framework of the EUROfusion Consortium, partially funded by the Euratom Research and Training Programme (Grant Agreement No. 101052200—EUROfusion) within the framework of the {\emph{Advanced Energetic Particle Transport models (ATEP)}} and {{\emph{TSVV-10}} projects.
The Swiss contribution to this work has been funded by the Swiss State Secretariat for Education, Research and Innovation (SERI).
Views and opinions expressed are however those of the authors only and do not necessarily reflect those of the European Union, the European Commission or SERI.
Neither the European Union nor the European Commission nor SERI can be held responsible for them.
Simulations were performed on the HPC-Marconi and HPC-M100 supercomputers.
% Part of this work has been done within the LABEX Plas@par project, and received financial state aid managed by the Agence Nationale de la Recherche, as part of the programme ``Investissements d'avenir'' under the reference ANR-11-IDEX-0004-02.
%This paper shows the results based on the invited presentation at the 46th EPS conference on Plasma Physics~\cite{BiancalaniEPS19}.

% 
% \section*{ORCID IDs}
% A Mishchenko: \url{https://orcid.org/0000-0003-1436-4502}\\
% T Hayward-Schneider: \url{https://orcid.org/0000-0003-0588-5090}\\
% E Poli: \url{https://orcid.org/0000-0001-7552-4800}
% R Kleiber  https://orcid.org/0000-0002-2261-2855
% L Villard  https://orcid.org/0000-0003-3807-9482

\end{document}